\documentclass[
reprint,
superscriptaddress,
amsmath,
amssymb,
aps,
prb,
floatfix,
]{revtex4-2}
\usepackage{graphicx}
\usepackage{hyperref}
\usepackage{xcolor}
\usepackage{multirow}

\begin{document}

\title{BraWl: Simulating the thermodynamics and phase stability of multicomponent alloys using conventional and enhanced sampling techniques}

\author{Hubert J. Naguszewski}
\affiliation{Department of Physics, University of Warwick, Coventry, CV4 7AL, United Kingdom}
\author{Livia B. P\'{a}rtay}
\affiliation{Department of Chemistry, University of Warwick, Coventry, CV4 7AL, United Kingdom}
\author{David Quigley}
\affiliation{Department of Physics, University of Warwick, Coventry, CV4 7AL, United Kingdom}
\author{Christopher D. Woodgate}
\email{christopher.woodgate@bristol.ac.uk}
\affiliation{H.H. Wills Physics Laboratory, University of Bristol, Royal Fort, Bristol, BS8 1TL, United Kingdom}

\begin{abstract}
We present BraWl, a Fortran package implementing a range of conventional and enhanced sampling algorithms for exploration of the phase space of the Bragg-Williams model, facilitating study of diffusional solid-solid transformations in binary and multicomponent alloys. These sampling algorithms include Metropolis-Hastings Monte Carlo, Wang-Landau sampling, and Nested Sampling. We demonstrate the capabilities of the package by applying it to some prototypical binary and multicomponent alloys, including high-entropy alloys
\end{abstract}


\date{November 3, 2025}

\maketitle

\section{Summary}
\label{sec:summary}

Many technologically relevant materials, both structural and functional, are `alloys'---systems in which two or more (typically) metallic elements are combined to produce a new material with desirable physical properties. In a substitutional alloy, there is a fixed underlying crystal lattice, while the probability of a given constituent element of the alloy occupying a particular lattice site is determined by thermodynamic considerations. In accordance with these considerations, atoms in an alloy can arrange themselves differently depending on the precise chemical composition and processing conditions. Frequently, a mixture of elements will form a regular crystalline lattice with substitutional disorder (a `solid solution') at high temperature, before atomic short- and long-range order emerges as the material is cooled. The nature of atomic arrangements in a material determines many important physical properties. For a given combination of elements, it is therefore crucial to understand the nature of atomic ordering in a material, as well as the temperature at which it emerges upon cooling, to guide materials processing strategies. One physically intuitive model for the internal energy of an alloy is the Bragg-Williams model, which assumes that atoms in the alloy interact in a pairwise manner. Crucially, the effective pair interactions (EPIs) which appear in the Bragg-Williams Hamiltonian can be obtained \textit{ab initio} using density functional theory (DFT) calculations. When appropriate sampling techniques are applied to the Bragg-Williams model, it is possible to explore the configuration space of a given alloy in detail and determine equilibrium phases as a function of temperature, leading to construction of phase diagrams. Here, we present \texttt{BraWl}, a Fortran package implementing a range of conventional and enhanced sampling algorithms for exploration of the phase space of the Bragg-Williams model, facilitating study of diffusional solid-solid transformations in binary and multicomponent alloys. These sampling algorithms include Metropolis-Hastings Monte Carlo, Wang-Landau sampling, and Nested Sampling. We demonstrate the capabilities of the package by applying it to some prototypical binary and multicomponent alloys, including high-entropy alloys.

\section{Statement of Need}
\label{sec:statement_of_need}

The Fortran package \texttt{BraWl} facilitates simulation of the thermodynamics and phase stability of both binary and multicomponent alloys. It achieves this by providing implementation of both the Bragg-Williams Hamiltonian (a lattice based model expressing the internal energy of an alloy as a sum of atom-atom effective pair interactions) concurrently with a range of conventional and enhanced sampling techniques for exploration of the alloy configuration space. The result is a package which can determine phase equilibria as a function of both temperature and alloy composition, which leads to the construction of alloy phase diagrams. Additionally, the package can be used for extraction of representative equilibrated atomic configurations for visualisation, as well as for use in complementary modelling approaches. Before outlining the specific need for this package, we first review some elementary details of alloy thermodynamics and statistical mechanics.

\subsection{Alloy thermodynamics}

In a substitutional alloy with fixed underlying lattice, a particular arrangement of atoms can be specified by a discrete set of site occupation numbers, $\left\{ \xi_{i\gamma} \right\}$, where $\xi_{i \gamma} = 1$ if site $i$ is occupied by an atom of species $\gamma$ and $\xi_{i \gamma} = 0$ otherwise. The lattice index $i$ takes values in the range 1 to $N$, where $N$ is the total number of lattice sites in the system, while the species index $\gamma$ takes values in the range 1 to $s$, where $s$ is the number of chemical species (elements) present in the alloy composition. The physical constraint that each lattice site is occupied by one (and only one)  atom is expressed as $\sum_{\gamma} \xi_{i\gamma} = 1$, while the total concentration of a given chemical species is given by $c_\gamma = \frac{1}{N} \sum_i \xi_{i\gamma}$. (Naturally, vacancies can be treated in this framework by considering them as an additional chemical species present at a very low concentration.) If we consider an ensemble of alloy configurations, we can then define the site-wise concentrations, $\left\{ c_{i\gamma}\right\}$, as the average value of the site occupation numbers across the ensemble, $c_{i\gamma} = \langle \xi_{i\gamma} \rangle$, where $\langle \cdot \rangle$ denotes the average taken over the ensemble. (The precise meaning of `ensemble' will be defined presently.)

\begin{figure*}[t]
    \centering
    \includegraphics[width=\textwidth]{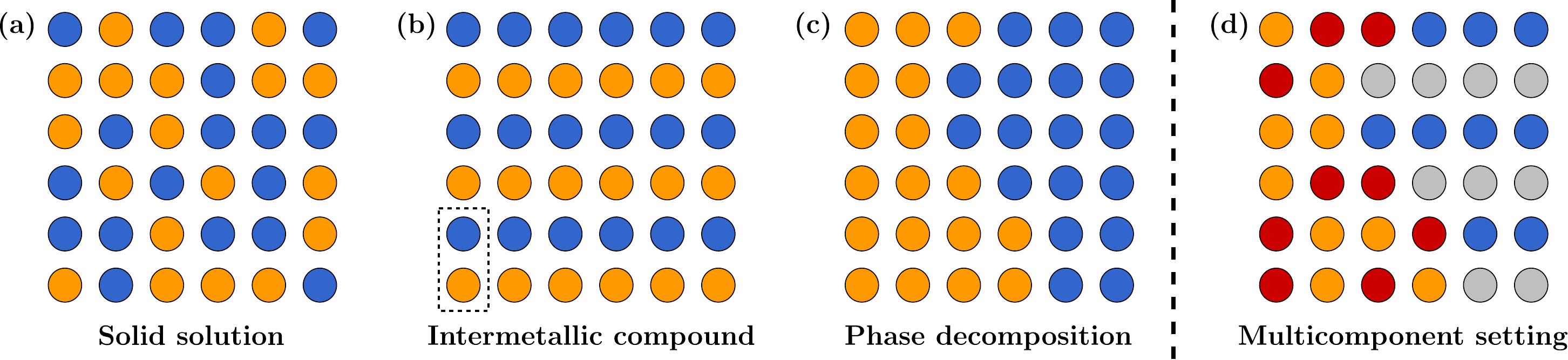}
    \caption{Illustrations of potential states of a substitutional alloy in thermal equilibrium. A solid solution (a) is a state where lattice sites are occupied at random by elements of different chemical species. An ordered intermetallic compound (b) has an identifiable regular, repeating motif of atoms. A system may also undergo phase decomposition (c), where pairs of elements phase segregate from one another. In the multicomponent setting (d) there can be many possible competing phases.}
    \label{fig:alloy_phase_cartoon}
\end{figure*}

In the context of substitutional alloys, the word \textit{phase} refers to a physically homogeneous area of the material with uniform chemical composition and physical properties. At the atomic level, this means that, on average, the site-wise concentrations in a particular spatial region are homogeneous and describe some repeating motif of atoms arranged on a lattice. In general, a particular single-phase region of an alloy can be either a `solid solution', in which all lattice sites have equal partial lattice site occupancies and elements mix randomly and homogeneously, or an ordered intermetallic phase (sometimes referred to as an `intermetallic compound'), where a crystallographically ordered, repeating motif of elements can be identified. Naturally, an intermetallic phase can also admit substitutional disorder on one or more of the atomic sites in the repeating motif. In thermal equilibrium, it is possible for more than one phase to be present in an alloy, with the maximum permitted number of phases present determined, in general, by the Gibbs phase rule. An alloy which decomposes into multiple coexisting phases in thermal equilibrium is said to undergo `phase segregation' or `phase decomposition'. (Sometimes this process can also be referred to as one or more phases `precipitating' out of the solid solution.) Some illustrations of potential alloy phases for a toy, 2D alloy in both binary and multicomponent settings, are provided in Fig.~\ref{fig:alloy_phase_cartoon}.

For a system with fixed underlying lattice, in the high temperature limit, entropy dictates that atoms should have no lattice site preference. However, as the temperature is lowered, this site symmetry will eventually be broken and either a partial or full site ordering (or decomposition) will become established. The relevant equilibrium phase(s) of an alloy are, in principle, fully determined once the pressure/volume, temperature, and alloy composition (\textit{i.e.} set of alloying elements and their concentrations) have been specified. Freezing the underlying crystal lattice removes the degree of freedom provided by the pressure/volume, and it remains to inspect the probabilities of a given arrangement of atoms occurring at the specified temperature. 

In thermal equilibrium, the relevant probability distribution for determining the likelihood of a given configuration occurring is the Boltzmann distribution. The partition function, $Z$, is written as
\begin{equation}
    Z = \sum_{\left\{ \xi_{i \gamma} \right\}} e^{-\beta E\left( \left\{ \xi_{i \gamma} \right\} \right)},
\end{equation}
where $E\left( \left\{ \xi_{i \gamma} \right\} \right)$ is the energy associated with a configuration, $\left\{ \xi_{i \gamma} \right\}$, and the sum is taken over all possible atomic configurations. The symbol $\beta$ is defined via $\beta=1/k_B T$, where $T$ is temperature, and $k_B$ is the usual Boltzmann constant. The probability of a given arrangement of atoms occurring is then
\begin{equation}
    P\left(\left\{ \xi_{i\gamma} \right\}\right) = \frac{e^{-\beta E\left( \left\{ \xi_{i \gamma} \right\} \right)}}{Z}.
\end{equation}
This probability distribution defines an ensemble of configurations distributed according to that probability. If the distribution is known, it is possible to recover various thermodynamic quantities of interest, such as the free energy and specific heat.

However, for any physically realistic form of the alloy internal energy, $E\left( \left\{ \xi_{i \gamma} \right\} \right)$, and for all but the smallest of simulation supercells, direct evaluation of the partition function is computationally intractable due to the huge number of configurations which must be considered. This is a particular problem in the context of alloys containing multiple elements, such as high-entropy alloys---those alloys containing four or more elements alloyed in near-equal ratios~\cite{george_high-entropy_2019}---as the size of the configuration space grows combinatorially both with the total number of atoms in the simulation cell, as well as with the number of elements present in a given composition~\cite{zhang_roadmap_2025}. Consequently, it is necessary both to find means by which to evaluate the energy associated with a given arrangement of atoms which are accurate and computationally efficient, as well as to use sampling algorithms to reliably estimate thermodynamic quantities and determine equilibrium phases as a function of temperature~\cite{ferrari_frontiers_2020, ferrari_simulating_2023} without evaluating the partition function in a brute-force manner. Evaluation of the internal energy of the alloy will be discussed now, while specific sampling algorithms (those which are implemented in \texttt{BraWl}) will be discussed later in this article.

\subsection{Evaluation of the alloy internal energy}

In the context of simulations performed on alloy supercells, the energy of a given configuration can be evaluated in a variety of ways~\cite{widom_modeling_2018}. These include first-principles electronic structure calculations using density functional theory (DFT)~\cite{eisenbach_first-principles_2019}; interatomic potentials, including machine-learned interatomic potentials (MLIPs)~\cite{rosenbrock_machine-learned_2021}; and lattice-based models such as cluster expansions (CEs)~\cite{ekborg-tanner_construction_2024}. All of these methods have their advantages and disadvantages. DFT calculations on alloy supercells, while highly accurate, are computationally expensive, rendering this option inviable when a large number of alloy energy evaluations are required for sampling. Calculations using interatomic potentials, MLIPs and CEs are significantly cheaper than direct DFT calculations, but they still frequently require a large DFT training dataset. Additionally, the number of fitting parameters required for these models grows significantly once a multicomponent alloy is considered, making them challenging to train and leading to concerns regarding their accuracy. Finally, it should be stressed that the computational cost of these models often remains prohibitive when a large number of energy evaluations is required.

An alternative, physically intuitive model for the internal energy of an alloy is the Bragg-Williams model~\cite{bragg_effect_1934, bragg_effect_1935}, which assumes that the internal energy of an alloy takes a simple, pairwise form. The Bragg-Williams {model} Hamiltonian has the form
\begin{equation}
    H(\{\xi_{i\gamma}\}) = \frac{1}{2}\sum_{i \gamma; j\gamma'} V_{i\gamma; j\gamma'} \xi_{i \gamma} \xi_{j \gamma'},
    \label{eq:b-w1}
\end{equation}
where $V_{i\gamma; j\gamma'}$ denotes the effective pair interaction (EPI) between an atom of chemical species $\gamma$ on lattice site $i$ and an atom of chemical species $\gamma'$ on lattice site $j$. (The factor of $\frac{1}{2}$ eliminates double-counting in the summation.) For a system of finite size, it is assumed that periodic boundary conditions are applied in all coordinate directions. We note that the form of this Hamiltonian is very similar to that of the Lenz-Ising model, an elementary model in magnetism~\cite{brush_history_1967}. 

Generally, the assumption is made that that the EPIs are spatially homogeneous and isotropic, and Eq.~\ref{eq:b-w1} is therefore rewritten as
\begin{equation}
    H(\{\xi_{i\gamma}\}) = \frac{1}{2}\sum_{i \gamma} \xi_{i \gamma} \left( \sum_{n} \sum_{j \in n(i)} \sum_{\gamma'} V^{(n)}_{\gamma \gamma'} \xi_{j \gamma'} \right),
    \label{eq:b-w3}
\end{equation}
where the sum over $i$ remains a sum over lattice sites, but the sum over $n$ denotes a sum over the coordination shells (nearest-neighbours, next-nearest-neighbours, \textit{etc.}) of the lattice. The notation $n(i)$ is then used to denote the set of lattice sites which are $n$th nearest-neighbours to site $i$. Then $V^{(n)}_{\gamma \gamma'}$ denotes the effective pair interaction between chemical species $\gamma$ and $\gamma'$ on coordination shell $n$. It is reasonable to assume that, for most alloys, the strength of EPIs will tail off quickly with decreasing distance, and the sum over $n$ can be taken over the first few coordination shells of the underlying lattice type being considered. (This is, of course, equivalent to imposing some radial `cutoff' on an interatomic potential.)

EPIs for the Bragg-Williams Hamiltonian can be obtained using a variety of methods, generally those based on density functional theory.  Similarly to the CE method, it is naturally possible to fit EPIs for a given alloy composition to a set of DFT total energy evaluations on alloy supercells with success~\cite{liu_machine_nodate, zhang_robust_2020, liu_monte_2021}. However, most frequently, such EPIs are obtained using the Korringa--Kohn--Rostoker (KKR) formulation of density functional theory~\cite{korringa_calculation_1947, kohn_solution_1954, ebert_calculating_2011}, where the coherent potential approximation (CPA) can be used to describe the average electronic structure and consequent internal energy of the disordered alloy~\cite{soven_coherent-potential_1967, gyorffy_coherent-potential_1972, stocks_complete_1978}. There are then a variety of suitable techniques available for assessing the energetic cost of applied, inhomogeneous chemical perturbations to the CPA reference medium which naturally lead to extraction of EPIs. Such techniques include both the generalised perturbation method (GPM)~\cite{ducastelle_generalized_1976, ruban_atomic_2004}, as well as techniques using the language of concentration waves to describe the atomic-scale chemical fluctuations~\cite{khachaturyan_ordering_1978, gyorffy_concentration_1983}. Approaches based on concentration waves have been derived for alloys both in the binary~\cite{staunton_compositional_1994, johnson_first-principles_1994} and multicomponent~\cite{singh_atomic_2015, khan_statistical_2016, woodgate_modelling_2024} settings. Once the EPIs for a given alloy composition are obtained, the phase stability of a particular alloy can be examined using sampling techniques applied to the Bragg-Williams model. This is the purpose of \texttt{BraWl} as presented in this work.

{
\subsection{Model limitations}

It should be emphasised that there are two key limitations to the above discussion, and to the applicability of the Bragg--Williams model more generally.

The first limitation is the lack of consideration of entropic contributions beyond that made by the configurational entropy, such as vibrational, magnetic, and electronic entropies. Perhaps most important is the consideration of vibrational entropy, which is most relevant when a system transitions between two phases with different elastic properties~\cite{van_de_walle_effect_2002}. For example, if (upon cooling) a system undergoes a phase transition from a disordered phase to an ordered phase, the latter of which is elastically stiffer than the former, the difference in vibrational entropy between the two phases can thermodynamically stabilise the disordered phase and lower the disorder-order transition temperature. Such effects are not accounted for in the Bragg-Williams model directly. However, they can be included in some approximate way by modifying EPIs to account for these effects. (One such scheme is discussed in Ref.~\cite{ruban_qualitative_2024}.) More generally, the effect of other entropic contributions, such as magnetic and electronic contributions, should also be considered as appropriate to a given system of study~\cite{ma_ab_2015}.

The second limitation is the assumption of a fixed underlying crystal lattice. In an alloy where there are atomic size mismatches, there will often be local distortions to the underlying crystal lattice to accommodate the mismatches. (This is because configurations with such local lattice distortions represent true minima of the potential energy surface.) It is understood that these effects can affect predicted disorder-order transition temperatures in alloys~\cite{kostiuchenko_impact_2019}. 

Neither of the above effects (additional entropic contributions or local lattice distortions) are explicitly considered in the fixed-lattice Bragg-Williams model as implemented in \texttt{BraWl}, and users should therefore take due care when interpreting simulation results.
}

\subsection{The purpose of \texttt{BraWl}}

There are a range of existing packages capable of simulating alloy phase equilibria, both open- and closed-source. Examples of widely-used such packages include ATAT~\cite{van_de_walle_alloy_2002}, ICET~\cite{angqvist_icet_2019} and CELL~\cite{rigamonti_cell_2024}, though all of these focus primarily on implementation of a general cluster expansion, rather than the simpler form of the Bragg-Williams Hamiltonian. To our knowledge, there is no open-source package specifically focussing on the implementation of a range of sampling algorithms applied to the Bragg-Williams model. We therefore believe that \texttt{BraWl} fills a gap in the capabilities of the current alloy software ecosystem. Additionally, we hope that the modular way in which the package is constructed could enable implementation of more complex Hamiltonians, as well as further sampling algorithms in addition to those detailed below, in due course.

\section{Sampling Algorithms}

\texttt{BraWl} implements a range of conventional and enhanced sampling algorithms for exploration of the alloy configuration space. At present, these are the Metropolis-Hastings algorithm, Wang-Landau sampling, and Nested Sampling. We briefly outline the details of each of these algorithms below.

\subsection{Metropolis--Hastings Monte Carlo}

The Metropolis--Hastings algorithm is a useful method for obtaining the equilibrium state of a system. It achieves this by allowing for a system of interest to follow a chain of states which evolve to, and sample, an equilibrium ensemble \cite{metropolis_equation_1953, landau_guide_2014}. For the Bragg-Williams model within the canonical ensemble, the algorithm functions by proposing a position swap between two randomly selected atoms in the simulation cell, and calculating the change in energy, $\Delta E$, induced by the swap. The swap is accepted with a probability given by
\begin{equation}
    P_{n\rightarrow m} = 
    \begin{cases}
        \; \text{exp}\left(-\Delta E/k_BT\right), &\quad \Delta E > 0\\
        \; 1, &\quad \Delta E \leq 0,
    \end{cases}
\label{eq:metropolis_transition_probability}
\end{equation}
where $n$ labels the initial state, $m$ labels the proposed (swapped) state, $T$ denotes the simulation temperature, and $k_B$ is the usual Boltzmann constant. These atom swaps can be performed according to Kawasaki dynamics \cite{kawasaki_diffusion_1966} (nearest neighbour swaps only) or performed between any two atoms in the system. (The latter option, while less physical, typically allows the system to reach equilibrium in fewer trial Monte Carlo moves.) Given enough trial moves, the system will reach equilibrium for a fixed simulation temperature. Once at equilibrium, decorrelated samples can be drawn to obtain thermodynamic averages of various quantities. Within \texttt{BraWl}, this functionality can be used to determine a range of quantities of interest, including atomic short- and long-range order parameters, as well as the simulation specific heat. (Definitions of these quantities are provided in Sec.~\ref{sec:physical_quantities}.) These quantities can be plotted as a function of temperature by considering simulations performed across a range of temperatures. It is also possible to perform `simulated annealing' where, starting at high temperature the system is equilibrated at a given temperature and statistics drawn, before the simulation temperature is decreased and the cycle repeated until a desired target temperature is reached. It also allows for determining the lowest available energy state of the alloy being studied to parametrise a Wang-Landau sampling, discussed below. Finally, for a given simulation temperature, it is possible to draw decorrelated atomic configurations which can then be used for visualisation and as inputs to other modelling techniques.

\subsection{Wang-Landau sampling}
\label{sec:wl}

Wang-Landau sampling is a flat histogram method which provides a means for high throughput calculation of phase diagrams for atomistic/lattice model systems \cite{wang_efficient_2001}. The method allows for direct computation of an estimate of the density of states in energy $g(E_i)$, and hence the partition function 
\begin{equation*}
Z = \sum_i g(E_i) e^{-\beta E_i},
\end{equation*}
where the index $i$ runs over the appropriately discretised energy macrostates of a given Hamiltonian.  Thermodynamic quantities at any temperature of interest can then be obtained provided one has prior knowledge of the minimum and maximum energy relevant to those temperatures. The method achieves this by starting from an initial `guess' of the density of states, which is used to generate a Markov chain of configurations which, as the guess is iteratively refined, tends toward a uniform sampling over energy. The uniformity is quantified by a  `flatness' criterion on a histogram of visited energies. The algorithm starts with making initially large modifications to the estimated density of states.  Once the flatness criterion is achieved, the modification factor is reduced and sampling begins again, a process which is repeated over multiple iterations until a desired tolerance on the modification factor is achieved.

Within the Bragg-Williams model, Wang-Landau sampling performs atom swaps as in the Metropolis--Hastings method but with the following acceptance criterion
\begin{equation}
    P_{n\rightarrow m} = 
    \begin{cases}
        \; \frac{g(E_n)}{g(E_m)}, &\quad g(E_n) < g(E_m)\\
        \; 1, &\quad g(E_n) \geq g(E_m),
    \end{cases}
\label{eq:wl_transition_probability}
\end{equation}
where $g$ is the density of states, $E_n$ is the initial energy and $E_m$ is the energy associated with the configuration where the proposed swap has been made. After each proposed swap, the density of states is updated according to
\begin{equation}
    g(E_i) \rightarrow g(E_i)f_k,
\end{equation}
where $E_i$ is the energy of the resultant state, $f_k$ is a modification factor initially ($k=0$) greater than 1, and $k$ is the current iteration index of the Wang-Landau sampling algorithm. A histogram of the energies visited is maintained, $H(E)$, as is a measure of the `flatness', $F$ of the histogram,
\begin{equation}
    F = \frac{\min(H(E))}{\frac{1}{N_b}\sum_i^N H(E_i)},
\end{equation}
where $N_b$ denotes the number of bins used in the histogram. Once $F$ is above a given tolerance, sampling is interrupted and $f$ is reduced for the next sampling iteration, \textit{e.g.} $f_{k+1}=\sqrt{f_k}$. The visit histogram $H(E)$ is set to zero and a new Wang-Landau iteration begins. This is repeated until $f$ falls within some desired tolerance, \textit{i.e.} sufficiently close to unity.

Within \texttt{BraWl}, the Wang-Landau sampling algorithm features a selection of parallelisation schemes using the message passing interface (MPI) as well as sampling enhancements. These schemes include energy domain decomposition with dynamic domain sizing and multiple random walkers per domain as well as replica exchange. This portion of \texttt{BraWl} is intended to be used to compute the simulation density of states (in energy) for a given alloy, from which a variety of data can be obtained such as energy distribution histograms at a given temperature as well as the specific heat across a desired temperature range.

\subsection{Nested Sampling}

Nested sampling (NS) is powerful Bayesian inference technique~\cite{skilling_nested_2004,skilling_nested_2006,ashton_nested_2022} adapted to sample the potential energy surface of atomistic systems, giving direct access to the partition function at arbitrary temperatures for comprehensive thermodynamic analysis, without relying on advance knowledge of relevant { atomic configurations} or the range of energies accessible to them~\cite{partay_efficient_2010, partay_nested_2021}.

Nested sampling is a top-down iterative approach, starting the sampling with random high-energy configurations and progressing towards the global minimum-energy structure through a series of consecutive nested energy levels. 
Within the Bragg-Williams model, these random high-energy configurations are configurations where the desired ratio of atomic species are assigned randomly to the available lattice sites. When the sampling is initialised, an integer number, $K$, of random alloy configurations are produced---these are often referred to as \emph{walkers} or the \emph{live set}---with $K$ controlling the resolution (and also the computational cost) of the sampling. 
During the iterative process, the configuration with the highest energy is recorded then removed from the live set, and substituted with a new configuration that has a lower energy, but uniformly randomly selected. The uniform distribution of the $K$ walkers allows the estimation of the relative phase space volume, $w_i= \{[K/(K+1)]^i-[K/(K+1)]^{i+1}\}$, at iteration $i$, and thus the partition function can be evaluated simply during a post-processing step at any arbitrary $\beta$,\cite{partay_nested_2021} as
\begin{equation}
    Z = \sum_{i} w_i~e^{-\beta E\left( \left\{ \xi_{i \gamma} \right\} \right)},
\end{equation}
where $E( \{ \xi_{i \gamma}\})$ is the energy of the configuration discarded in the $i$th NS iteration. Thermodynamic quantities and weighted average of observables can be evaluated from this.
Since a simple rejection sampling quickly becomes unaffordable as lower energy regions need to be sampled, the new configurations are generated by a random walk in practice: one of the existing $K$ samples is selected randomly, cloned, and a series of species swap moves are proposed between randomly selected lattice sites, accepting every swap unless it would cause the energy to increase above the limit. However, as swap moves cannot be adjusted as the sampling progresses (unlike \textit{e.g.} atomic displacement steps), the acceptance ratio of swaps decreases. Thus, to ensure that new configurations are different from the starting structure, the number of proposed particle swap moves is doubled each time the acceptance probability (\textit{i.e.} the ratio of the number of atoms moved during a set of swaps compared to the total number of atoms in the simulation cell) falls below 5\%.

In this work, since the lattice sites are fixed and the Hamiltonian is discretised, it is possible to create multiple different configurations with numerically the same energy value. However, as the NS algorithm must be able to select the unique highest energy configuration during the iterative sampling process, we have to avoid such degeneracy. Thus, the energy of each configuration is perturbed by a positive, uniform random number of value less than $10^{-8}$~Ry, making each energy value numerically distinct without effecting the uniform distribution of samples or any thermodynamic properties.

\section{Physical Quantities of Interest}
\label{sec:physical_quantities}

\texttt{BraWl} can extract a range of quantities of interest from a given alloy simulation. The relevant quantities are:

\subsection{Internal energy} 

For a given lattice type, system size, alloy composition, and set of atom-atom effective pair interactions, \texttt{BraWl} can evaluate the total energy associated with the alloy configuration (Eq.~\ref{eq:b-w1}). At the time of writing, for speed, common lattice types (fcc, bcc, simple cubic, \textit{etc.}) are hard-coded, with the intention that the range of implemented lattice types will be expanded over time as necessary. Where only the relative \textit{change} of energy induced by swapping/substituting atoms is considered, \texttt{BraWl} takes advantage of the mathematical form of the Bragg-Williams Hamiltonian to only evaluate the relevant terms in the summation in Eq.~\ref{eq:b-w1} which are changed as a result of the swap/substitution. This substantially reduces the cost of evaluation of change in simulation energy induced by a particular atomic swap.

\subsection{Specific heat} 

The isochoric (fixed volume) specific heat at a given temperature, $C_V(T)$, is a useful quantity for identifying phase transitions, as a plot of the simulation specific heat as a function of temperature is expected to show a local peak at the temperature at which the transition occurs. Within \texttt{BraWl}, the specific heat is calculated via
\begin{equation}
    C_V(T) = \frac{\langle E^2\rangle - \langle E\rangle^2}{k_BT^2},
\end{equation}
where $k_B$ is the usual Boltzmann constant, $E$ is the simulation energy, and $\langle \cdot \rangle$ denote thermodynamic averages obtained using the relevant sampling algorithm.

\subsection{Atomic short-range order (ASRO) parameters} 

To assess local atom-atom correlations in a simulation, \texttt{BraWl} can calculate the Warren-Cowley atomic short-range order parameters~\cite{cowley_approximate_1950, cowley_short-range_1965}, adapted to the multicomponent setting, defined as
\begin{equation}
    \alpha^{\gamma \gamma'}_n=1-\frac{P^{\gamma \gamma'}_n}{c_{\gamma'}}
\end{equation}
where $n$ refers to the $n$\textsuperscript{th} coordination shell, $P^{\gamma \gamma'}_n$ is the conditional probability of an atom of type $q$ neighbouring an atom of type $p$ on shell $n$, and $c_q$ is the total concentration of atom type $q$. When $\alpha^{\gamma \gamma'}_n > 0$, $p$-$q$ pairs are disfavoured on shell $n$ and, when $\alpha^{\gamma \gamma'}_n < 0$ they are favoured. The value $\alpha^{\gamma \gamma'}_n = 0$ corresponds to the ideal, maximally disordered solid solution.

For maximal flexibility, \texttt{BraWl} outputs the conditional probabilities, $P^{\gamma \gamma'}_n$, and the user can then choose whether or not to rescale and obtain the Warren-Cowley ASRO parameters as a post-processing step. The \texttt{BraWl} package can calculate these parameters averaged across a single configuration, or averaged at a particular temperature using one of the sampling algorithms implemented in the code.

\subsection{Atomic long-range order (ALRO) parameters} 

Over a simulation run (for example using the Metropolis algorithm), \texttt{BraWl} can calculate the average partial occupancies of a lattice site, $\langle \xi_{i\gamma}\rangle = c_{i\gamma}$, which are natural atomic long-range order parameters describing chemically ordered phases. This capability was first demonstrated on simulations of Fe-Ga alloys~\cite{marchant_ab_2021}.

\section{Example Applications}

\begin{figure}[t]
\centering
\includegraphics[width=\linewidth]{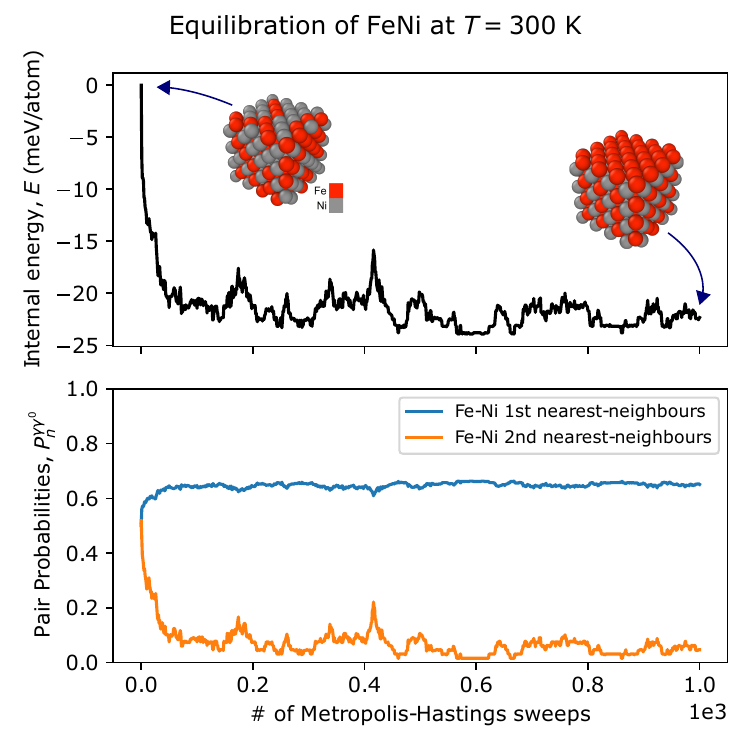}
\caption{Evolution of the simulation internal energy (top panel) and conditional pair probabilities (bottom panel) for an Fe$_{0.5}$Ni$_{0.5}$ alloy as a function of the number of Metropolis-Hastings sweeps at a simulation temperature of $T=300$~K. One `sweep' is one trial move per atom in the system. Beyond approximately 100 sweeps, the system can be seen to have reached equilibrium, with L1$_0$ order established.}
\label{fig:feni_equilibration}
\end{figure}

\texttt{BraWL} has been used, with success, to study the phase behaviour of a range of binary and multicomponent alloys, for example the binary Fe-Ga system (Galfenol)~\cite{marchant_ab_2021}, the Fe-Ni system~\cite{woodgate_integrated_2024}, the Cantor-Wu medium- and high-entropy alloys~\cite{woodgate_compositional_2022, woodgate_interplay_2023}, the refractory high-entropy alloys~\cite{woodgate_short-range_2023, woodgate_competition_2024}, the Al\textsubscript{$x$}CrFeCoNi system~\cite{woodgate_structure_2024}, and the AlTiVNb and AlTiCrMo refractory high-entropy superalloys~\cite{woodgate_emergent_2025}. Additionally, the package has been used to generate atomic configurations {with physically motivated ASRO and/or ALRO for subsequent study using a range of other simulation techniques. We highlight examples of its use in generating a training dataset for a machine-learned interatomic potential for the prototypical austenitic stainless steel, Fe$_7$Cr$_2$Ni~\cite{shenoy_collinear-spin_2024}, as well as its use in generating configurations for use in a transition state study for ferromagnetic Fe-Ni alloys~\cite{fisher_lattice_nodate}. Finally, the package has also been used to benchmark the efficiency of various parallelisation strategies proposed for the Wang-Landau sampling algorithm~\cite{naguszewski_optimal_nodate}. In this work, we explicitly consider several illustrative examples of the results which can be obtained using the sampling algorithms outlined above applied to the Bragg-Williams model as implemented in the package.}

As an example of the Metropolis-Hastings Monte Carlo algorithm, we consider its application to the binary FeNi alloy, first discussed in Ref.~\cite{woodgate_integrated_2024}. Fig.~\ref{fig:feni_equilibration} shows the internal energy and conditional pair probabilities (quantifying ASRO) of a simulation cell containing 256 atoms as a function of the number of Metropolis-Hastings `sweeps', where a sweep refers to performing a number of trial Metropolis-Hastings moves equal to the number of atoms in the simulation cell. The simulation is performed at 300~K, below the alloy's L1$_0$ disorder-order transition temperature. The L1$_0$ phase is a structure where $2/3$ of the nearest neighbours of Fe (Ni) atoms are Ni (Fe) atoms, and where none of the next-nearest neighbours of Fe (Ni) atoms are Ni (Fe) atoms. It can be seen that this ordering is swiftly established as the number of Monte Carlo sweeps increases, albeit with some remaining thermal noise.

\begin{figure}[t]
\centering
\includegraphics[width=\linewidth]{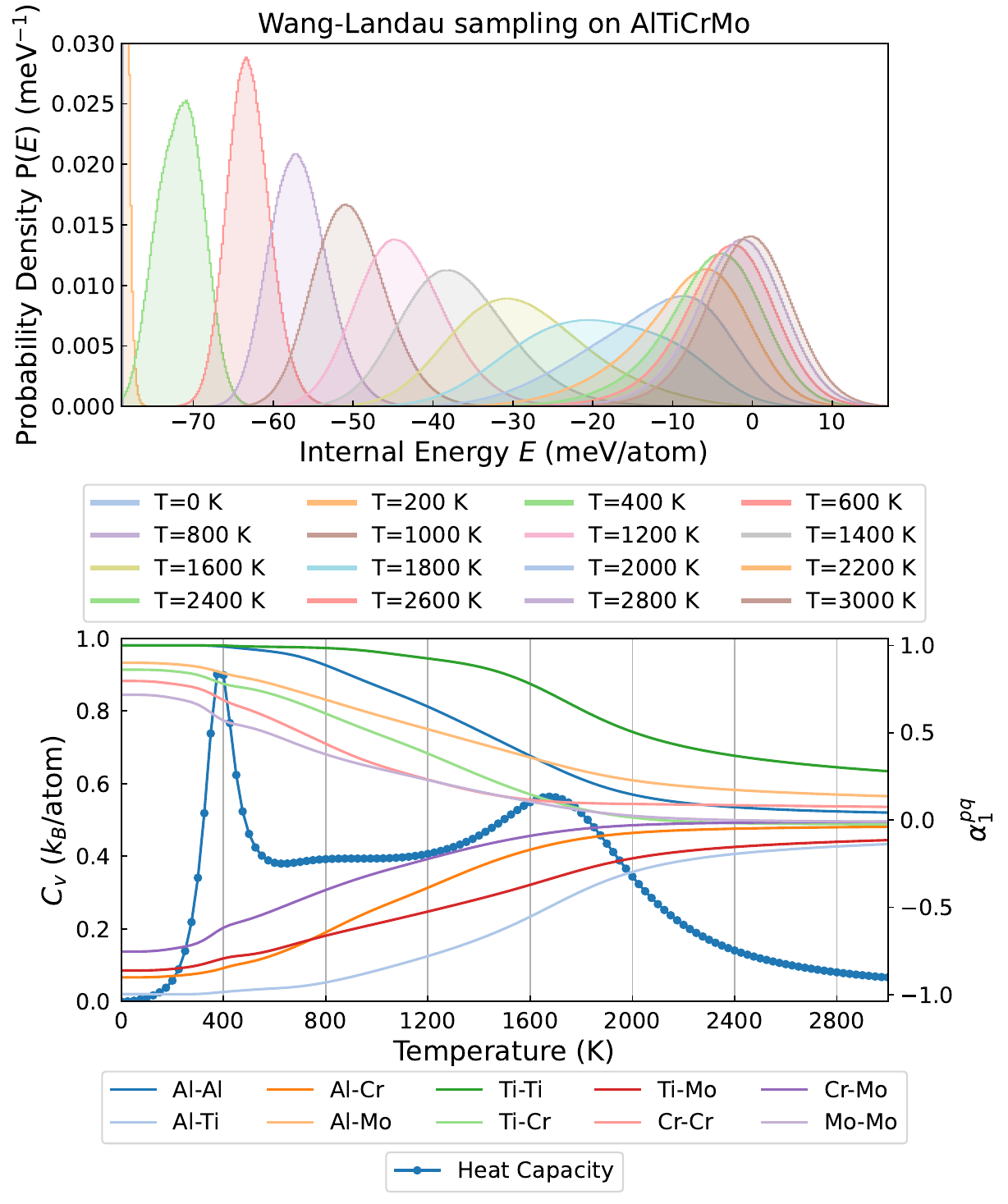}
\caption{Plots of energy probability distributions, Warren-Cowley ASRO parameters ($\alpha^{pq}_n$) and simulation specific heat ($C_V$) as a function of temperature for AlTiCrMo obtained using lattice-based Monte Carlo simulations employing Wang-Landau sampling. Here, show $\alpha^{pq}_n$ only for $n$ = 1. The zero of the energy scale for the energy histograms is set to be equal to the average internal energy of the alloy obtained at a simulation temperature of 3000~K.}
\label{fig:wl_AlTiCrMo}
\end{figure}

As an example of Wang-Landau sampling, we consider its application to the AlTiCrMo refractory high-entropy superalloy, first discussed in Ref.~\cite{woodgate_emergent_2025}, for which results are shown in Fig.~\ref{fig:wl_AlTiCrMo}. The top panel shows calculated energy probability distributions (histograms) at various temperatures, while the bottom panel shows the simulation specific heat and evolution of the Warren-Cowley ASRO parameters as a function of temperature. The high-temperature peak in the specific heat data is associated with the experimentally observed B2 crystallographic ordering.

\begin{figure}[t]
\centering
\includegraphics[width=\linewidth]{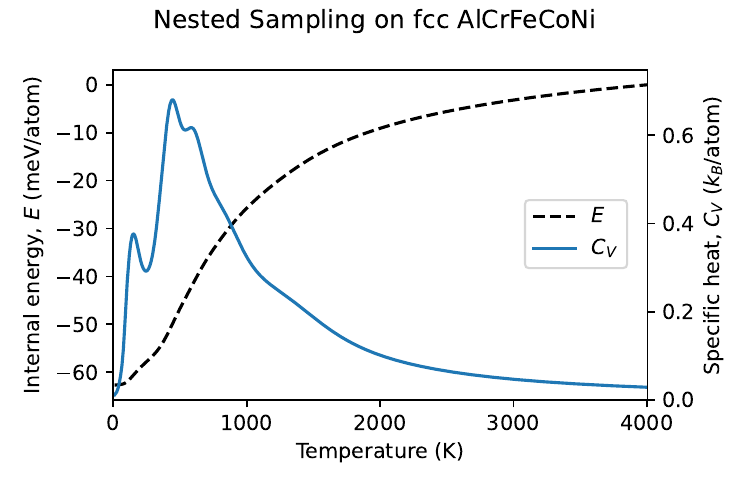}
\caption{Internal energy, $E$, and isochoric specific heat, $C_V$, obtained using the Nested Sampling algorithm applied to the equiatomic, fcc, AlCrFeCoNi high-entropy alloy. The simulation cell contained 108 atoms. Upon cooling, the initial peak in the specific heat is associated with an L1$_2$ ordering driven by Al, with subsequent peaks indicating eventual decomposition into multiple competing phases.}
\label{fig:ns_alcrfeconi}
\end{figure}

Finally, as an example of application of the Nested Sampling algorithm, we consider its application to the AlCrFeCoNi high-entropy alloy, first discussed in Ref.~\cite{woodgate_structure_2024}. Fig.~\ref{fig:ns_alcrfeconi} plots the internal energy, $E$, and isochoric specific heat, $C_V$, obtained for the equiatomic, fcc, AlCrFeCoNi system. The simulation cell contained 108 atoms. The initial peak in the specific heat encountered upon cooling from high temperature is associated with an L1$_2$ ordering driven by Al, with subsequent peaks indicating eventual decomposition into multiple competing phases, which is understood to be consistent with experimental data for this system.

{
\section{Performance}

Fundamentally, all of the outlined sampling algorithms---Metropolis--Hastings Monte Carlo, Wang--Landau sampling, and Nested Sampling---require a large number of evaluations of the alloy internal energy (Eq.~\ref{eq:b-w1}) during the sampling process. The time taken for evaluation of the model Hamiltonian is therefore the key factor in determining the computational cost of a sampling run on a given system.

In the context of the present work, two key observations should be made regarding how the computational cost of a simulation scales with the size of a simulation cell containing $N$ atoms. The first observation relates to how evaluation of the Bragg--Williams model Hamiltonian scales with respect to system size. As the atom-atom EPIs are assumed to be of finite range and each atom therefore only interacts with other atoms within some finite cutoff radius (c.f. Eq.~\ref{eq:b-w3}), evaluation of the model Hamiltonian scales approximately linearly with $N$ for systems which are sufficiently large. The second relates to how the number of configurations accessible to a simulation cell scales with $N$. Formally, for an $s$ species alloy with $N_1$ atoms of species 1, $N_2$ atoms of species 2, and so on, the number of possible atomic arrangements of these atoms on the $N$ available lattice sites, $n_\textrm{configs}$, is
\begin{equation}
    n_\textrm{configs} = \frac{N!}{N_1!\times N_2! \times \dots \times N_s!}.
\end{equation}
For an equiatomic binary alloy ($N_1 = N_2 = N/2$) this means that for a 32-atom cell there are approximately $6.01 \times 10^{8}$ possible configurations, while for a 256-atom cell this number becomes approximately $5.79 \times 10^{75}$. For an equatomic quarternary system ($N_1 = N_2 = N_3 = N_4 = N/4$), these numbers become approximately $7.93 \times 10^{29}$ and $3.31 \times 10^{150}$ respectively. Accordingly, for large systems with many elements in the alloy composition, longer sampling runs are anticipated to be needed to ensure the configuration space has been adequately explored and for any calculated quantities to be well-converged. We emphasise that, naturally, any given sampling algorithm is not anticipated to visit all possible system configurations, only a representative example that is converged.

To assess the performance of the \texttt{BraWl} package, we measure the time taken for a Metropolis--Hastings run at a single temperature, as this quantifies the rate of sampling and thus the performance of all of the considered algorithms. As a benchmark, we consider a simulation on the equiatomic AlTiVNb alloy using the EPIs of Ref.~\cite{woodgate_emergent_2025}---which include interactions up to and including sixth-nearest neighbours---and a simulation cell consisting of $8\times8\times8$ bcc unit cells with $N=1024$ in total. The sampling run consisted of a Metropolis-Hastings run with 1,024,000 trial Metropolis--Hastings swaps ($10^3$ trial moves per atom) during an initial burn-in phase, followed by a sampling run of 10,240,000 trial atomic swaps ($10^4$ trial moves per atom) during which statistics are gathered. The energy of the simulation was stored every 1,024 trial moves for a total of 10,000 decorrelated energy samples, while the Warren-Cowley parameters up to and including the second coordination shell of the lattice were evaluated and stored every 10,240 trial moves for a total of 1,000 decorrelated ASRO samples. (Typically, fewer samples are required to reliably converge ASRO results than are required to estimate quantities such as as the specific heat.) For a simulation performed locally in serial on an Apple M3 Pro CPU (ARM architecture) with \texttt{BraWl} built using the GNU compiler collection \texttt{v15.1.0} at \texttt{-O3} compiler optimisation, this simulation run took an average of 12.5~s across three attempts, indicating an average of approximately 900,000 trial atomic swaps per second. On an 4.9~GHz 12th generation Intel Core i7-12700 CPU (x86 architecture) with \texttt{BraWl} built using the 2022 Intel Fortran compiler at \texttt{-O2} compiler optimisation, this simulation run took an average 31.8~s across three attempts, indicating an average of approximately 350,000 trial atomic swaps per second.
}

\section*{acknowledgments}
H.J.N. and C.D.W. acknowledge invaluable training in scientific software development provided by Dr C. S. Brady and Dr H. Ratcliffe in their Introduction to Scientific Software Development course at the University of Warwick.  L.B.P. and C.D.W. acknowledge useful discussions with Dr Georgia A. Marchant (Department of Physics and Astronomy, Uppsala University, Sweden). Finally, C.D.W. acknowledges guidance from Prof. Julie B. Staunton (Department of Physics, University of Warwick, UK) during the early stages of this work.


H.J.N. was supported by a studentship within the Centre for Doctoral Training in Modelling of Heterogeneous Systems, funded by the UK Engineering and Physical Sciences Research Council (EPSRC), Grant EP/S022848/1. L.B.P. acknowledges support from the EPSRC through the individual Early Career Fellowship, Grant EP/T000163/1. C.D.W. acknowledges support from an EPSRC Doctoral Prize Fellowship at the University of Bristol, Grant EP/W524414/1. 

Computing facilities for development and testing of the package were provided by the \href{https://www.bristol.ac.uk/acrc/}{Advanced Computing Research Centre (ACRC)} of the University of Bristol, and by the \href{https://warwick.ac.uk/research/rtp/sc/}{Scientific Computing Research Technology Platform (SCRTP)} of the University of Warwick. We also acknowledge use of the Sulis Tier 2 HPC platform. Sulis is funded by EPSRC Grant EP/T022108/1 and the HPC Midlands+ consortium.

\section*{Code and Data Availability}

\texttt{BraWl} is fully open-source and is available freely under the GNU Lesser General Public License (\texttt{LGPL-3.0}) at: \href{https://github.com/ChrisWoodgate/BraWl}{https://github.com/ChrisWoodgate/BraWl}. Scripts for generating relevant data and reproducing the plots visualised in this manuscript are available in the \texttt{examples/} subdirectory of this repository. The code is also available on Zenodo: \href{https://doi.org/10.5281/zenodo.10371917}{https://doi.org/10.5281/zenodo.10371917}.

\end{document}